\documentstyle[aps,12pt,multicol,amssymb]{revtex}
\input{epsf.tex}
\newcommand{\SS}{\scriptscriptstyle}
\newcommand{\beq}{\begin{equation}}
\newcommand{\eeq}{\end{equation}}
\newcommand{\bea}{\begin{eqnarray}}
\newcommand{\eea}{\end{eqnarray}}
\begin{document}

\title{Uptake of gases in bundles of carbon nanotubes}

\author {George Stan$^{1,3,6}$, Mary J. Bojan$^2$, Stefano Curtarolo$^{1,4}$,\\
 Silvina M. Gatica$^{1,5}$, and Milton W. Cole$^{1}$}
\address{ $^1$Department of Physics, Penn State University, University
Park, PA 16802, USA \\
$^2$Department of Chemistry, Penn State University, University
Park, PA 16802, USA \\
$^3$Present address: Institute for Physical Science $\&$ Technology and\\ 
Department of Chemical Engineering, University of Maryland, College Park, 
MD 20742, USA\\
$^4$Present address: Department of Materials Science and Engineering,\\ 
MIT, Cambridge, MA 02139, USA\\ 
$^5$Permananent address: Departamento de F\'{\i}sica, Facultad de Ciencias Exactas y Naturales,\\ 
Universidad de Buenos Aires, Buenos Aires, Argentina.\\
$^6${corresponding author, e-mail: gns@ipst.umd.edu, phone: (301) 405-4812, 
fax: (301) 314-9404}}
\date{\today}

\maketitle
\begin{abstract}

Model calculations are presented which predict whether or not an arbitrary
gas experiences significant absorption within carbon nanotubes and/or
bundles of nanotubes. The potentials used in these calculations assume a
conventional form, based on a sum of two-body interactions with individual
carbon atoms; the latter employ energy and distance parameters which are
derived from empirical combining rules. The results confirm intuitive
expectation that small atoms and molecules are absorbed within both the
interstitial channels and  the tubes, while large atoms and molecules are
absorbed almost exclusively within the tubes.
\end{abstract}

\baselineskip 22pt

\section{Introduction}

The absorption of gases in nanopores is a subject of growing
experimental and theoretical interest, stimulated by both fundamental
scientific questions and the potential for many technologies \cite{dillon,rodriguez,inoue,teizer,nutzenadel,orimo,migone,migone1,yates,ye,chen,stan1,stan2,stan3,radhakrishnan,wang1,mays,vidales,levesque,liu,dresselhaus}. One of the most 
important questions to be addressed is whether or not a specific gas is 
significantly absorbed within carbon nanotubes; we will define the word 
``significant'' in Eq. \ref{eq:rho} below. While the answer depends in detail 
on the specific thermodynamic conditions of the coexisting vapor (pressure $P$ 
and temperature $T$), one expects that intuitive considerations based on size 
and energy scales ought to provide useful qualitative insights. For example, it
 has been demonstrated  that gases whose condensed phases possess low surface 
tensions are strongly imbibed in these tubes \cite{dujardin}. This important 
result can be understood from either the Kelvin equation or a comparison of 
competing interaction (adhesive vs. cohesive) energies. These considerations 
arise in the analogous problem of wetting transitions \cite{cheng}.

This paper addresses this basic question by employing a simple, but plausible, 
model of the interaction potential, from which we compute the adsorption as a 
function of $P$ and $T$. We assume that the adsorption potential can be derived
 from a sum of Lennard-Jones (LJ) two-body interactions between the host C 
atoms and the adsorbate. This pair potential has distance and energy parameters
 obtained with semiempirical combining rules from the  LJ $\epsilon$ and 
$\sigma$ parameters of the C atoms and the adsorbate \cite{steele1,steele2,scoles,ihm}:
\bea
\sigma_{g{\SS C}} &=& {\sigma_{gg}+\sigma_{\SS CC}} \over 2\\
\epsilon_{g{\SS C}} &=& \sqrt{\epsilon_{gg}\epsilon_{\SS CC}} \nonumber
\eea
where ``g'' and ``C'' refer to the gas and C atoms, respectively.
 Estimates of the gas parameters  are given for some relevant systems in
Table \ref{table:LJ} \cite{watts,maitland,klein}, while for C atoms we use
$\sigma_{\SS CC} = 3.4$ \AA\ and $\epsilon_{\SS CC} = 28$ K \cite{steele2}. 
These values are typical, but uncertain within a 15\% range \cite{bruch}.

This paper's approach is the following. We first choose a particular
(somewhat arbitrary) criterion for calling the uptake ``significant''. For
example, Fig. \ref{fig:10}(a) and most of our work employ the criterion 
\beq
\rho^*=\rho \sigma_{g{\SS C}} = 0.1
\label{eq:rho}
\eeq
where $\rho=N/L$ is the one-dimensional (1d) density, with $N$ the number of 
adsorbed atoms and $L$ the length of the tube, and $\rho^*$ is the 
corresponding dimensionless density. For gases of interest here, this 
criterion corresponds to a mean 1d spacing of order 30 \AA. This is a very 
low density. Although we do consider more stringent criteria elsewhere in 
this paper, the results do not differ qualitatively. The reason for the lack 
of sensitivity to the threshold is that once adsorption commences, it rises 
rapidly as a function of $P$ (until the crowding effect of repulsive forces 
slows the variation of coverage with $P$).

In the nanotube bundle geometry, adsorption can take place inside the tubes,
in the interstitial channels, and on the outer surface of the nanotube bundle 
(Fig. \ref{fig:sites}). Typical length scales for the triangular lattice of 
nanotubes in the bundle are: lattice constant 17 \AA, nanotube radius 
6.9 \AA, bundle diameter between 50 and 100 \AA, and bundle length 
$\sim 10-100\ \mu m$ \cite{smalley}. We will see that size is a critical 
variable determining uptake. Some key findings of this paper appear in 
Fig. \ref{fig:10}(a), which shows the uptake at a very small ratio of $P$ to 
saturated vapor pressure ($P_0$). Small atoms and molecules (which typically 
have small values of $\epsilon_{gg}$) are strongly adsorbed within both the 
nanotubes and the narrow interstitial channels (IC's) between nanotubes. 
Larger particles, in contrast, do not ``fit'' within the IC's but do imbibe 
within the tubes. Perhaps a surprising feature of Fig. \ref{fig:10}(a) is that 
a hypothetical gas with a very large value of $\epsilon_{gg}$ adsorbs in 
neither place. This occurs because the relative tendency (compared with bulk 
condensation) of a gas to be absorbed within the tube {\it at a given 
 undersaturation} depends on the ratio of adhesive to cohesive energies. 
The geometric mean combining rule for $\epsilon_{g{\SS C}}$ implies that this 
ratio varies as the inverse square root of $\epsilon_{gg}$, so a large 
$\epsilon_{gg}$ implies small uptake. This finding is qualitatively consistent 
with the empirical correlation between uptake and surface tension mentioned 
above. It also correlates with the physics determining wetting behavior of 
liquids for which the analogous comparison involves the same kind of 
interaction ratio \cite{cheng}.

This paper makes a number of simplifications in order to draw such
general conclusions. Arguably the most drastic assumptions are that the
nanotubes are infinite and perfect and that the nanotube bundles involve a
unique species of tubes in a regular array (geometry unaffected by the 
adsorption). 

The outline of this paper is the following. Section \ref{sec:pot} describes 
our model of the interactions. Section \ref{sec:stat} presents the statistical 
mechanical model used in the calculations. Section \ref{sec:results} reports 
our results. Section \ref{sec:concl} summarizes these and discusses open 
questions.

\section{Adsorption potential}
\label{sec:pot}

A basic assumption in our model is that the potential energy
experienced by a molecule at position ${\bf r}$ can be evaluated by a 
summation of two-body interactions $U ({\bf x})$ between the molecule and the 
carbon atoms comprising the tube: 
\beq
V({\bf r}) = \sum_i U ({\bf r}-{\bf R_i})
\eeq
This assumption is made in the overwhelming majority of calculations of gas
interactions with either graphite and carbon nanotubes. In the graphite
case, many-body effects have been found to be $\sim$ 15\% corrections 
to {\it ab initio} pair potential sums \cite{kim}. Hence, the empirical pair
potential should be regarded as an effective pair potential. One might expect 
somewhat smaller many body contributions in the nanotube case 
because the molecule is somewhat farther from the nearest carbon atom 
\cite{distance} and because the effective coordination number is larger in the 
nanotube case than on graphite. In contrast, the argument in the IC case leads 
to the prediction of a larger many body effect than on graphite. 
These expectations, however, might {\it not} be correct because the many-body 
expansion involves geometry-dependent competing terms of opposite signs 
\cite{kim1} and because the two body energy for the IC is typically of much 
larger magnitude than on a flat surface.

Another key assumption made here is that the pair potential is isotropic
and of LJ form: $U(x) = 4\epsilon [(\sigma/x)^{12}-(\sigma/x)^6]$.
There is {\it ab initio} and empirical evidence to the effect that anisotropy 
of the pair potential plays a role in adsorption potentials on graphite
\cite{carlos}. Nevertheless, most studies of adsorption on that surface 
neglect such an effect and use a LJ pair potential similar to what we use here.
 The  final assumption is the use of an azimuthally and longitudinally 
averaged potential. The potential at distance $r$ from the axis of the 
cylinder is then \cite{stan1}:
\beq
V(r;R)=3 \pi\ \theta\ \epsilon\ \sigma^2\  \biggl[ \frac{21}{32} \biggl(\frac{\sigma}{R}\biggr)^{10} f_{11}(x) M_{11}(x) - \biggl( \frac{\sigma}{R}\biggr)^4 f_5(x) M_5(x) \biggr]
\eeq
where $\theta$ = 0.38 \AA$^{-2}$ is the surface density of C atoms and $R$ is 
the radius of the cylinder. Here, $x=r_</r_>$, and $r_{<(>)}$ are the smaller
(greater) of $r$ and $R$. The function $f_n (x)$ is defined as 1 for $r < R$ 
and $(R/r)^n$ for $r > R$, with $n$ a positive integer. Here we use the 
integrals
\begin{equation}
M_{n}(x) = \int_0^{\pi} d\varphi \frac{1}{(1 + x^2 - 2 x cos\varphi)^{n/2}}
\end{equation}
We emphasize that each approximation introduces an error, but the qualitative 
trends ought to be reliable. At this time, the lack of high quality {\em ab
initio} calculations would seem to warrant this kind of approach.

The IC potential is obtained by summing up the contribution from three 
nanotubes and azimuthally averaging the result. Figures 
\ref{fig:NT_contour}-\ref{fig:GR_contour} show contour plots in the 
$\sigma_{gg}-\epsilon_{gg}$ plane of the reduced minimum of the adsorption 
potential ($V_{min}^* \equiv V_{min}/\epsilon_{gg}$) for all of these sites. 
Inside both the tubes and in IC's there is a threshold $\sigma_{gg}$ value 
above which the potential becomes repulsive, corresponding to gases which are 
too big to fit in these restricted geometries; these thresholds are 
$\sigma_{gg} \simeq $ 11.4 \AA\ (tubes) and 3.4 \AA\ (IC's) for nanotubes of 
radius 6.9 \AA\ studied here. Outside of the bundle, there are no such size 
constraints for the adsorbed atoms/molecules, as the adsorbate can always 
find a region in which the potential is attractive; at a fixed value of 
$\epsilon_{gg}$ large systems yield larger $|V_{min}^*|$ due to their larger 
coordination number of C atoms. In all three cases the most negative values of 
$V_{min}^*$ occur for small values of $\epsilon_{gg}$. In the tube and external
 surface cases, but not the IC case, the most negative values of $V_{min}^*$ 
occur for large $\sigma_{gg}$ ($\gtrsim 9$ \AA).

\section{Statistical Mechanics}
\label{sec:stat}

Our interest is whether atoms are likely to go inside the tubes, in the 
interstitial channels, and on the outer surface of the nanotube bundle. This
behavior is determined by the thermodynamic conditions ($P$, $T$) and 
microscopic parameters (especially $\sigma_{gg}$ relative to $R$). A key 
factor implicit here is the cohesive energy of the bulk phase of the adsorbate 
which determines a relevant pressure, i. e. saturated vapor pressure $P_0$. We 
construct a simple model for the low coverage regime of atoms inside nanotubes,
 neglecting the interactions between adsorbate atoms, while for atoms moving 
in the very confining IC's any density can be considered because of the 
mathematical simplicity resulting from the 1d character of the system. We have 
discussed elsewhere the extreme quantum behavior of He at low 
T \cite{stan3,cole}. In the present case we assume that classical statistical 
mechanics applies \cite{lowT}.
 
We now compute the chemical potential $\mu$ of the adsorbate. All of our 
calculations take the coexisting three-dimensional vapor to be an ideal gas, 
so that the chemical potential can be expressed in terms of pressure
 as $\mu = \beta^{-1} ln(\beta P \lambda^3)$. Here $\beta^{-1}= k_B T$, and 
$\lambda = (2 \pi \hbar^2 \beta/m)^{1/2}$ is the de Broglie thermal wavelength
for particles of mass $m$. It is convenient to measure the chemical 
potential with respect to its value at saturation, $\mu_0$,
\beq
\Delta \mu = \mu - \mu_0 = \beta^{-1} ln(P/P_0)
\eeq
An analytical expression for $P_0$ is available from computer simulation data 
of the Lennard-Jones system's liquid-vapor coexistence \cite{lotfi}, 
$ln P^*_0 = 1.2629 T^* - 4.9095/T^* - 0.15115/T^{*4}$, where 
$P^*_0 = P_0 \sigma^3_{gg}/\epsilon_{gg}$ and $T^* = k_B T/\epsilon_{gg}$
are reduced quantities.

Consider first the adsorption inside a single nanotube. The chemical potential 
of the ideal gas in an external potential can be expressed as a function of 
the number of adsorbed atoms, $N$, and temperature:
\beq
e^{\beta \mu} = {{N \lambda^3} \over {\int_{\SS NT} d{\bf r}\ exp(-\beta V(r))}}
\eeq
where the integral 
is performed over the volume of the nanotube. This is an application of Henry's
 law. Then, the chemical potential relative to its value at saturation assumes 
a simple form due to the cylindrical symmetry of the adsorption potential:
\beq
\Delta \mu = \beta^{-1} ln \left({\rho \over {2 \pi \beta P_0 
\int_{\SS NT} dr\ r\ exp(- \beta V(r))}}\right)
\eeq

Atoms in the narrow IC's are strongly confined to the vicinity of the axis so 
that a 1d model is applicable and solvable for all densities. As 
previously discussed in the case of very small nanotubes \cite{stan1}, the 
transverse motion may be treated independently of the longitudinal motion and 
the chemical potential in this case has the form:
\beq
\mu = \mu_{\SS \perp} + \mu_{\SS 1d}
\label{eq:mu}
\eeq
where $\mu_{\SS \perp}$ is the transverse contribution and $\mu_{\SS 1d}$ is 
the chemical potential of a 1d gas. In general, $\beta \mu_{\SS \perp} = ln (\sum_i exp(-\beta \epsilon_i))$, where $\{\epsilon_i\}_{i=0,1,...}$ is the 
transverse spectrum of individual atoms/molecules. At low T ($\beta (\epsilon_1 - \epsilon_0) << 1 $), the ground state dominates the sum and 
$\mu_{\SS \perp} \simeq \epsilon_0$. The ground state energy can be determined 
very accurately using the WKB method \cite{brack}, since the adsorption 
potential is well-approximated by a parabola in the vicinity of the IC axis 
\cite{stan3}. Results for the potential well depths of various gases are shown 
in Table \ref{table:LJ}. 

The 1d chemical potential is obtained by integrating the 1d Gibbs-Duhem 
relation
\beq
{\partial \mu_{\SS 1d} \over \partial P_{\SS 1d}} = {1 \over \rho}
\label{eq:Gibbs-Duhem}
\eeq
where $P_{\SS 1d}$ is the 1d pressure. The particle density in the case of 
only nearest-neighbor interactions is given by the equation of state 
\cite{bojan,takahashi}
\beq
\rho = \frac{\int_0^{\infty} dz \ exp(- \beta [u(z) + z P_{\SS 1d}])}{\int_0^{\infty} dz \ z \ exp(- \beta [u(z) + z P_{\SS 1d}]) } 
\label{eq:rho_1d}
\eeq
Here u(z) is the LJ potential describing the interactions between adsorbed 
atoms. The integration of eq. \ref{eq:Gibbs-Duhem} leads to 
\beq
\beta \mu_{\SS 1d} = ln \left( \frac{\beta \lambda P_{1d,0} \int_0^{\infty} dz \ exp(- \beta [u(z) + z P_{1d,0}]) }{\int_0^{\infty} dz \ exp(- \beta [u(z) + z P_{\SS 1d}])} \right)
\label{eq:mu_1d}
\eeq
$P_{1d,0}$ is an initial low pressure chosen such that the ideal gas limit is 
reproduced. The density dependence of the 1d chemical potential is finally 
obtained by eliminating the 1d pressure between eqs. \ref{eq:Gibbs-Duhem}, 
\ref{eq:rho_1d} and \ref{eq:mu_1d}.

As shown in Fig. \ref{fig:GR_contour}, the external surface of the nanotube 
bundle also provides an attractive domain of adsorption. We have studied 
adsorption in the very attractive groove-like channel which runs  parallel to 
the nanotube axes, as shown in Figure \ref{fig:sites} \cite{grooves}. Then, a 
procedure similar to that employed in the case of IC's is applicable to 
computing the coverage. The contribution of the longitudinal motion to the 
chemical potential is determined in the same fashion as in the case of IC's. 
The ground state energy of the transverse motion can be estimated through a 
parabolic approximation for the adsorption potential at this site. Values of 
the ground state energy ($E_0^{ext}$) obtained in this fashion for the systems 
studied, as well as the well-depth of the adsorption potential 
($V_{min}^{ext}$), are listed in Table \ref{table:LJ}.

\section{Results}
\label{sec:results}

The lines obtained by setting the coverage equal to the threshold criterion
can be seen in Figures \ref{fig:10}(a)-(d). Figure \ref{fig:10}(a) shows this 
behavior in the case of $\rho^* = 0.1$, for 
$\Delta \mu^* = \Delta \mu/\epsilon_{gg} = -10$ and $T^* = 1$. As expected, 
small atoms or molecules (He, Ne, H$_2$) fit easily inside both the tubes and 
IC's, while large molecules do not fit in the narrow IC's. Hypothetical 
(but nonexistent) atoms with $\sigma_{gg} < 2.5 $ \AA\ are adsorbed in the 
IC's only if their self-interaction energy ($\epsilon_{gg}$) does not exceed a 
threshold value. The upper limit to the molecular size for adsorption 
inside the tubes can be seen in Fig. \ref{fig:10.high}. Indeed, the 
experimental observation of C$_{60}$ molecules encapsulated in nanotubes 
\cite{smith} is consistent with this expectation (as the point near 
$\sigma_{gg} \sim 9$ \AA\ indicates). 

Including the effect of interactions does not affect our results significantly.
In the framework of the gas-surface virial expansion \cite{steele},
\beq
N \simeq e^{\beta \mu} Q_1 [1+ \rho\ \eta (T)]
\eeq
where $Q_1$ is the single particle canonical partition function, and 
\beq
\eta (T) = L\ \frac{\int_{NT} d {\bf r_1}\ d{\bf r_2}\ exp(-\beta\ (V (r_1) + V (r_2))\ [e^{-\beta u (|{\bf r_1 -r_2}|) }-1]}{[\int_{NT} d {\bf r}\ exp(-\beta\ V (r))]^2}
\eeq
The net effect of the virial correction is at most a 0.1\% change of 
$\Delta \mu$; such a small magnitude is consistent with the expected behavior 
in the low pressure regime of interest here.

The evolution of the diagram as a function of the adsorption criterion can be 
seen in Fig. \ref{fig:10}(b). As the threshold density decreases 
($\rho^* = 0.05$ here) more systems satisfy the uptake criterion. 
Fig. \ref{fig:10}(c) shows a similar effect on the diagram of an increase in
chemical potential, to $\Delta \mu^* = -8$. In both geometries, the altered
criterion corresponds to more systems being allowed in the respective cavities.
A different effect on the diagram occurs if the size of the nanotubes is 
changed, as shown in Fig. \ref{fig:10}(d), under the thermodynamic conditions 
of Fig. \ref{fig:10}(a). In the case of nanotubes with radius 8 \AA\ more atoms
 enter interstitial channels because of the larger channel space, while fewer 
atoms go inside the tubes because the adhesive energy decreases. The 
trends seen in Fig. \ref{fig:10}(a)-(d) are qualitatively consistent with the 
behavior of $V_{min}^*$ presented in Figures \ref{fig:NT_contour} and 
\ref{fig:IC_contour}.
 
In the Henry's law regime of low coverage there is a convenient way to 
characterize the variation of uptake with geometry. We compute the ratio of 
particle occupations in the nanotubes and IC's at the same $P$ and $T$:
\beq
\Gamma (\epsilon_{gg}, \sigma_{gg}) = \frac{\nu_{\SS NT}}{\nu_{\SS IC}} \frac{\int_{\SS NT} d{\bf r}\ exp(-\beta V)}{\int_{\SS IC} d{\bf r}\ exp(-\beta V)}
\eeq
where $\nu_{\SS NT(IC)}$ are the number of nanotubes (IC's) in the bundle and 
the integrations are over one region (assumed infinitely long). For an 
infinite array of nanotubes, $\nu_{\SS NT}/\nu_{\SS IC} = 1/2$. The finiteness 
of the bundle changes the ratio; however, there is no qualitative effect on 
our conclusions unless the bundle is very small. This ratio depends on the two 
gas parameters, $\epsilon_{gg}$ and $\sigma_{gg}$. In order to simplify the 
presentation, we fit the general trend of systems in Table \ref{table:LJ} to 
an empirical equation:
\beq
\epsilon_{gg}^{fit} \simeq a\ \sigma_{gg} + b
\label{eq:fit}
\eeq
with values $a = 147$ K/\AA\ and $b = 376$ K. We then consider a function of 
one variable 
\beq
\Gamma (\sigma_{gg}) \equiv \Gamma (\epsilon_{gg}^{fit}, \sigma_{gg})
\eeq
This ratio function is presented in Figures \ref{fig:henry} and 
\ref{fig:henry.77}. In Fig. \ref{fig:henry} we consider a common value of 
$T^* = 1$, while in Fig.\ref{fig:henry.77}, we consider a fixed $T = 77$ K. 
The data in Fig. \ref{fig:henry} shows, as expected, that large (small) 
molecules adsorb preferentially in the nanotubes (IC's). 
Fig. \ref{fig:henry.77} differs for small $\sigma_{gg}$ because at 77 K the 
entropic advantage of the tubes is manifested as a larger uptake there than is 
seen in Fig. \ref{fig:henry} at the much lower T given by Eq. (\ref{eq:fit}).

\section{Conclusions}
\label{sec:concl}

Model calculations were used to investigate adsorption in nanotube bundles.
Simplifying assumptions were made, such as the pairwise summation of 
gas-surface interactions, the use of combining rules to determine energy and 
size parameters, and the continuum, rigid model of the carbon atoms of 
the tube. We studied mainly the regime of low coverage, where interactions 
between adsorbed atoms are omitted; in the IC case, this assumption was not 
needed, as a quasi-one dimensional approximation permits exact treatment of LJ 
interactions at finite coverages. The conclusions drawn are expected to be 
qualitatively accurate in general situations, and so they provide useful 
insight for experiments. The key result appears in Fig. \ref{fig:10}, 
indicating which molecules go where under ``typical'' experimental conditions. 
More general behavior can be estimated from the reduced potential curves
(Figures \ref{fig:NT_contour}-\ref{fig:GR_contour}) we have presented.

Williams and Eklund \cite{williams} have computed the H$_2$ adsorption on 
the bounding surface of bundles containing a finite number of tubes. In some 
cases, this contribution can be a significant fraction of the total adsorption.
Adsorption isotherms of classical gases on the external surface of the bundle 
is the subject of our current investigations to be reported in the future.

We discuss the relevant experiments very briefly. Teizer {\it et al.} 
\cite{teizer} studied He uptake and found consistency with our calculations
for one-dimensional motion and the computed binding energy within the 
interstitial channels. Kuznetsova {\it et al.} \cite{yates} studied uptake of 
Xe and their data are consistent with our calculations of the uptake within the
 nanotubes.

Interestingly, a recent experimental study of adsorption of methane in
nanotube bundles \cite{migone1} concluded that significant IC adsorption
occurs. This conclusion was reached from the fact that nanotubes were capped
and the measured binding energy of CH$_4$ determined (2570 K) was 
76\% larger than that on graphite (1460 K) \cite{vidali}, which compared 
favorably with previous estimates of the IC binding energy of H$_2$, He and 
Ne \cite{stan1,stan2,stan3}. Our present calculations indicate, however, that 
the large size of CH$_4$ prevent it from populating the narrow IC's 
significantly. In contrast, the external surface of the nanotube bundle is 
accessible and the binding energy in this case (Table \ref{table:LJ}) is 
$\sim $ 20\% larger than the one for graphite. A 
more realistic potential exhibits corrugation, which we have neglected here;
its effect is to increase the binding energy \cite{stan2}, but we have not 
undertaken that calculation as yet so no definite comparison is possible. 

\acknowledgments
We are grateful to Victor Bakaev, Moses Chan, Vincent Crespi, Peter Eklund, 
Karl Johnson, James Kurtz, Aldo Migone, Bill Steele and Keith Williams for 
useful discussions. This research was supported by the National Science 
Foundation, the Petroleum Research Fund of the American Chemical Society and 
the Army Research Office. One of the authors, S. C., would like to acknowledge 
the generous support of {\it Fondazione Ing. Aldo Gini}.

\newpage

\begin{table}
\center
\caption{The values of the LJ parameters, $\epsilon_{gg}$ and $\sigma_{gg}$, 
for the gas-gas interactions, the corresponding minimum ($V_{min}$)
 of the adsorption potential and ground state energies ($E_0$) inside a 
nanotube ($NT$), in the interstitial channel ($IC$), on the external surface 
of the bundle ($ext$), and on a single graphite sheet ($GR$) 
is given. LJ parameters were taken from Ref. \protect\cite{watts}, except for 
CH$_4$, CF$_4$ and SF$_6$ \protect\cite{maitland} and C$_{60}$ 
\protect\cite{klein}.}

\vspace*{0.2 in}
\begin{tabular}{ccccccccc} 
Gas &$\epsilon_{gg}$ (K)& $\sigma_{gg}$ (\AA)&$V_{min}^{\SS IC}$ (K)&$E_{0}^{\SS IC}$ (K)&$V_{min}^{\SS NT}$ (K)& $V_{min}^{ext}$ (K)& $E_0^{ext}$ (K)& $V_{min}^{GR}$ (K)\\ \hline
He      & 10.2 & 2.56  &-546  &   -386&  -297& -367&  -270& -218 \\
Ne      & 35.6 & 2.75  &-1018 &   -902&  -600& -725&  -666& -431 \\
H$_2$   & 37.0 & 3.05  &-828  &   -292&  -690& -808&  -618& -482 \\
Ar      & 120  & 3.40  &6     &    228& -1426&-1607& -1550& -965 \\
CH$_4$  & 148  & 3.45  &401   &    789& -1614&-1809& -1714&-1088 \\
Kr      & 171  & 3.60  &2048  &   2250& -1836&-2025& -1981&-1220 \\
Xe      & 221  & 4.10  &14786 &  15054& -2523&-2617& -2580&-1593 \\
CF$_4$  & 157  & 4.58  &36411 &  36854& -2539&-2475& -2433&-1520 \\
SF$_6$  & 208  & 5.25  &136492& 137196& -3726&-3307& -3272&-2056 \\
C$_{60}$&2300  & 9.2   &52858932&52863770&-49071&-21952&-21924&-14505\\
\end{tabular}
\label{table:LJ}
\end{table}

\newpage
\begin{center}
{\bf Figure Captions}
\end{center}

\begin{figure}
\caption{ (a) Diagram indicating regions of significant uptake 
(at thermodynamic conditions specified by $\Delta \mu^* = -10$, $T^* = 1$ 
and $\rho^* = 0.1$) as a function of the adsorbate Lennard-Jones parameters.
Gases lying in the domain deoted ``TUBE'' are absorbed within the nanotubes. 
Those denoted ``IC'' are absorbed within the interstitial channels, while those
denoted ``BOTH'' (``NEITHER'') go to both places (neither place). 
Systems of particular interest are identified by dots ($\bullet$), with 
parameters listed in Table \ref{table:LJ}.\\ 
(b) Diagram analogous to (a), except that curves shown utilize an alternative 
$\rho^*$ value for the threshold condition, i.e. $\rho^*= 0.05 $.\\ 
(c) Same as in (a) for a different value of the chemical potential: 
$\Delta \mu^* = -8$.\\
(d) Same as in (a) in the case of a nanotube array with tubes of diameter 16 
\AA.}
\label{fig:10}
\end{figure}

\begin{figure}
\caption{ Schematic picture of adsorption sites within and outside
a nanotube bundle. For the external surface, the most attractive site, 
located at equal distance from two nanotubes, is shown here. Adsorbed atoms or 
molecules are represented by dots.}
\label{fig:sites}
\end{figure}

\begin{figure}
\caption{Contour plot of the reduced well-depth $V_{min}^*$ of the adsorption 
potential inside a carbon nanotube. The attractive isopotential curves (---) 
correspond to $V_{min}^*$ increments of 10 starting from -90, while the 
repulsive curves (--- ---), from left to right, correspond to $V_{min}^* = $ 
20, 40, and 80.}
\label{fig:NT_contour}
\end{figure}

\begin{figure}
\caption{Same as in Fig. \ref{fig:NT_contour}, for the interstitial channel. 
Repulsive curves, from left to right, correspond to $V_{min}^* =$ 10, 20, 30 
and 40.}
\label{fig:IC_contour}
\end{figure}

\begin{figure}
\caption{Same as in Fig. \ref{fig:NT_contour}, for the external surface of the 
nanotube bundle.}
\label{fig:GR_contour}
\end{figure}

\begin{figure}
\caption{Expanded version of Fig. \ref{fig:10}(a) showing the gas systems
which absorb within a nanotube at $\rho^* = 0.1$, $\Delta \mu^* = -10$ and 
$T^* =1$.}
\label{fig:10.high}
\end{figure}

\begin{figure}
\caption{Ratio of the amount adsorbed inside a nanotube to that in an 
intersititial channel in the Henry's law (low coverage) regime. Here, 
$T^* =1$.}
\label{fig:henry}
\end{figure}

\begin{figure}
\caption{Same as in Fig. \ref{fig:henry}, but for $T = 77$ K.}
\label{fig:henry.77}
\end{figure}

\newpage
\begin{figure}[ht]
\epsfysize=5.in \epsfbox{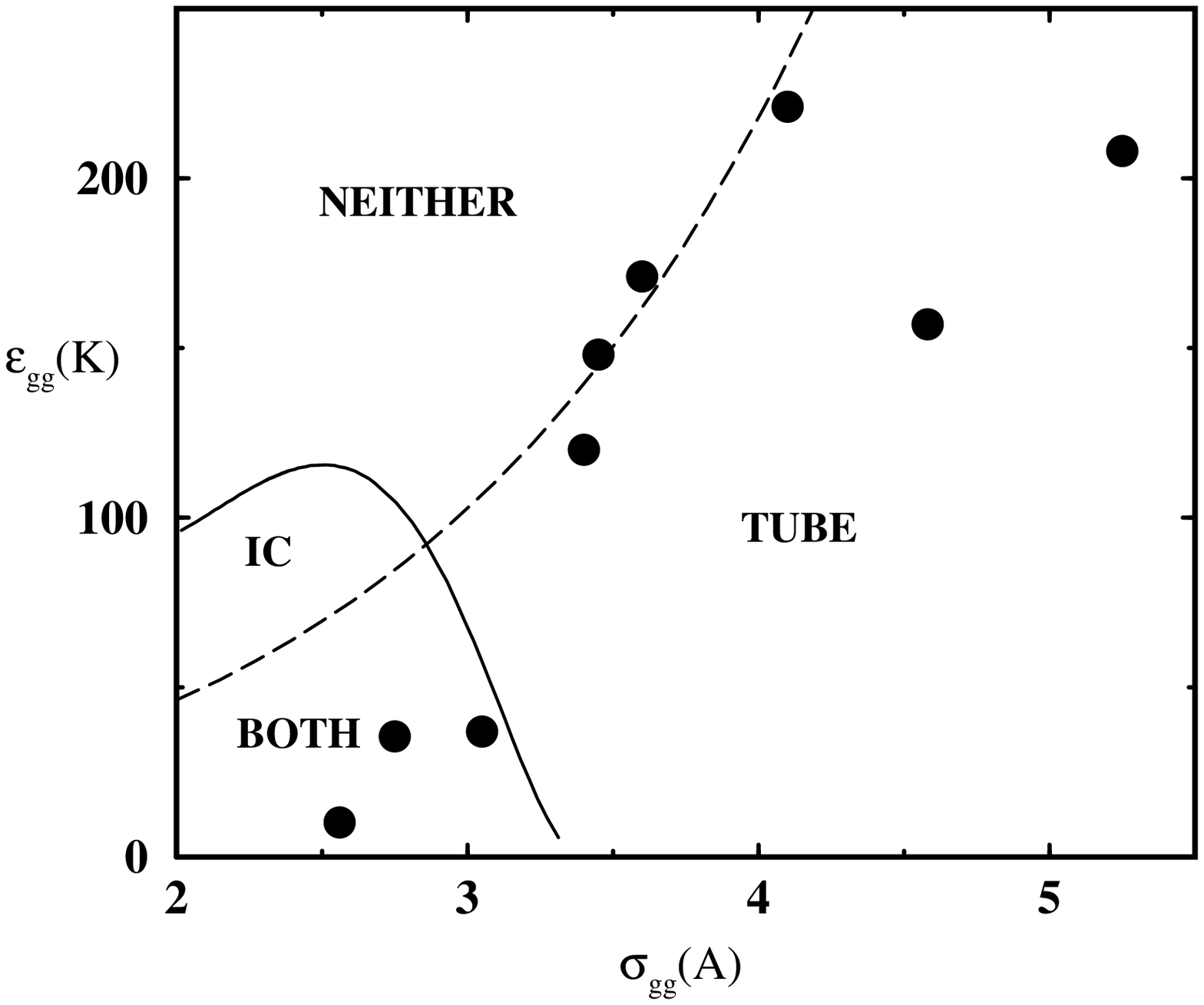}
\end{figure}
\vspace{7cm}
\begin{center}
{\bf FIG. \ref{fig:10}(a)}
\end{center}

\newpage
\begin{figure}[ht]
\epsfysize=5.in \epsfbox{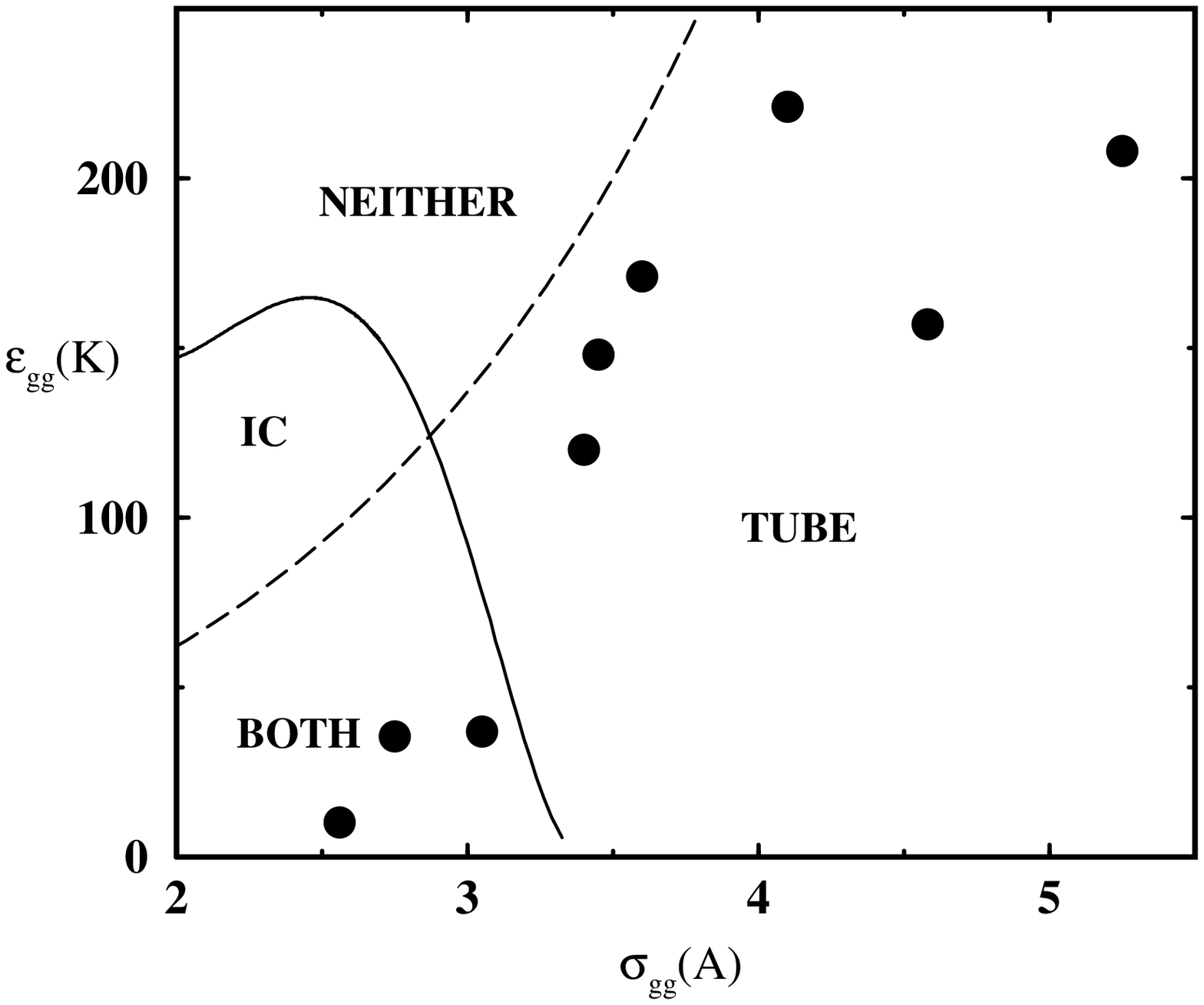}
\end{figure}
\vspace{7cm}
\begin{center}
{\bf FIG. \ref{fig:10}(b)}
\end{center}

\newpage
\begin{figure}[ht]
\epsfysize=5.in \epsfbox{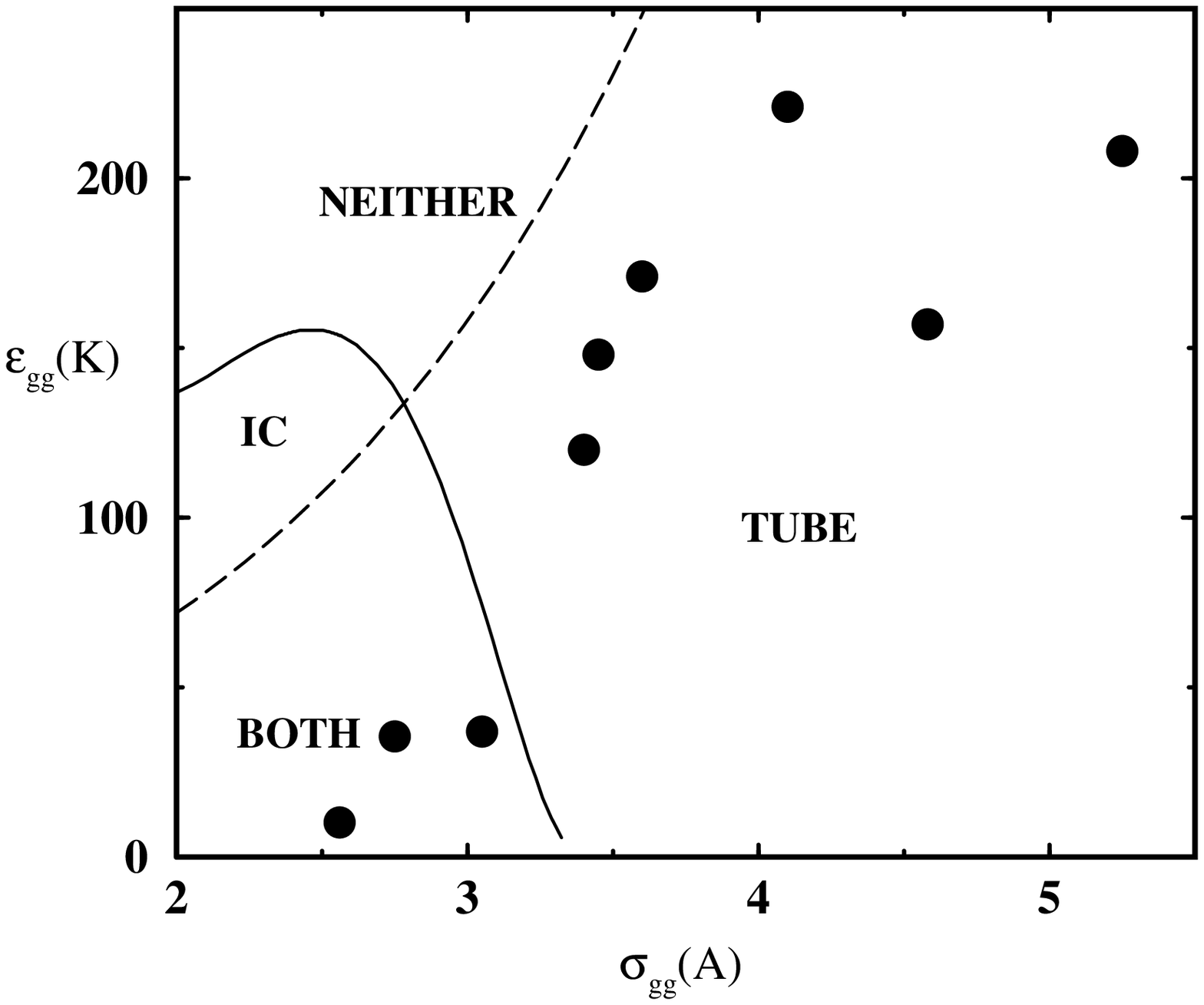}
\end{figure}
\vspace{7cm}
\begin{center}
{\bf FIG. \ref{fig:10}(c)}
\end{center}

\newpage
\begin{figure}[ht]
\epsfysize=5.in \epsfbox{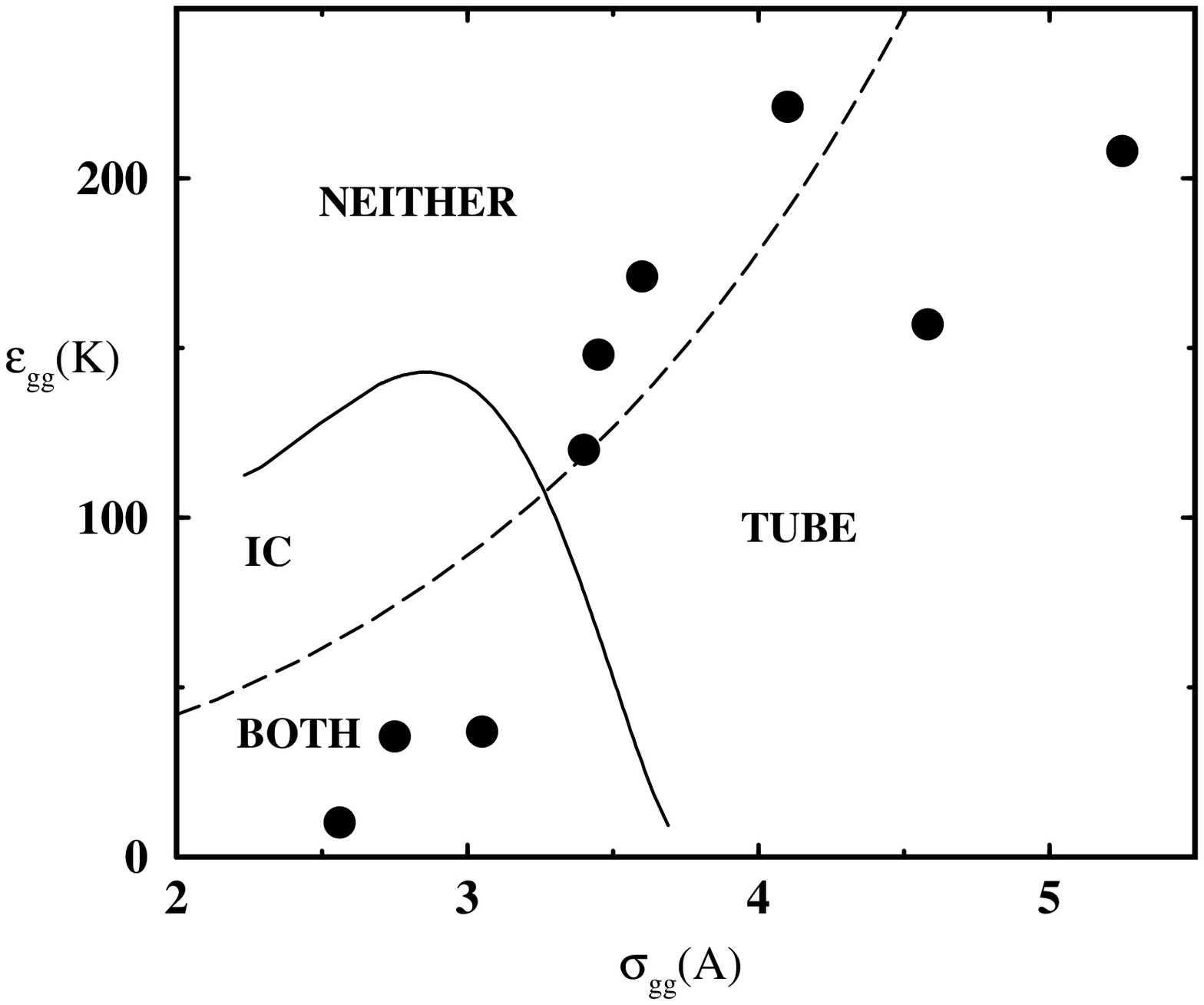}
\end{figure}
\vspace{7cm}
\begin{center}
{\bf FIG. \ref{fig:10}(d)}
\end{center}

\newpage
\begin{figure}[ht]
\epsfysize=4.8in \epsfbox{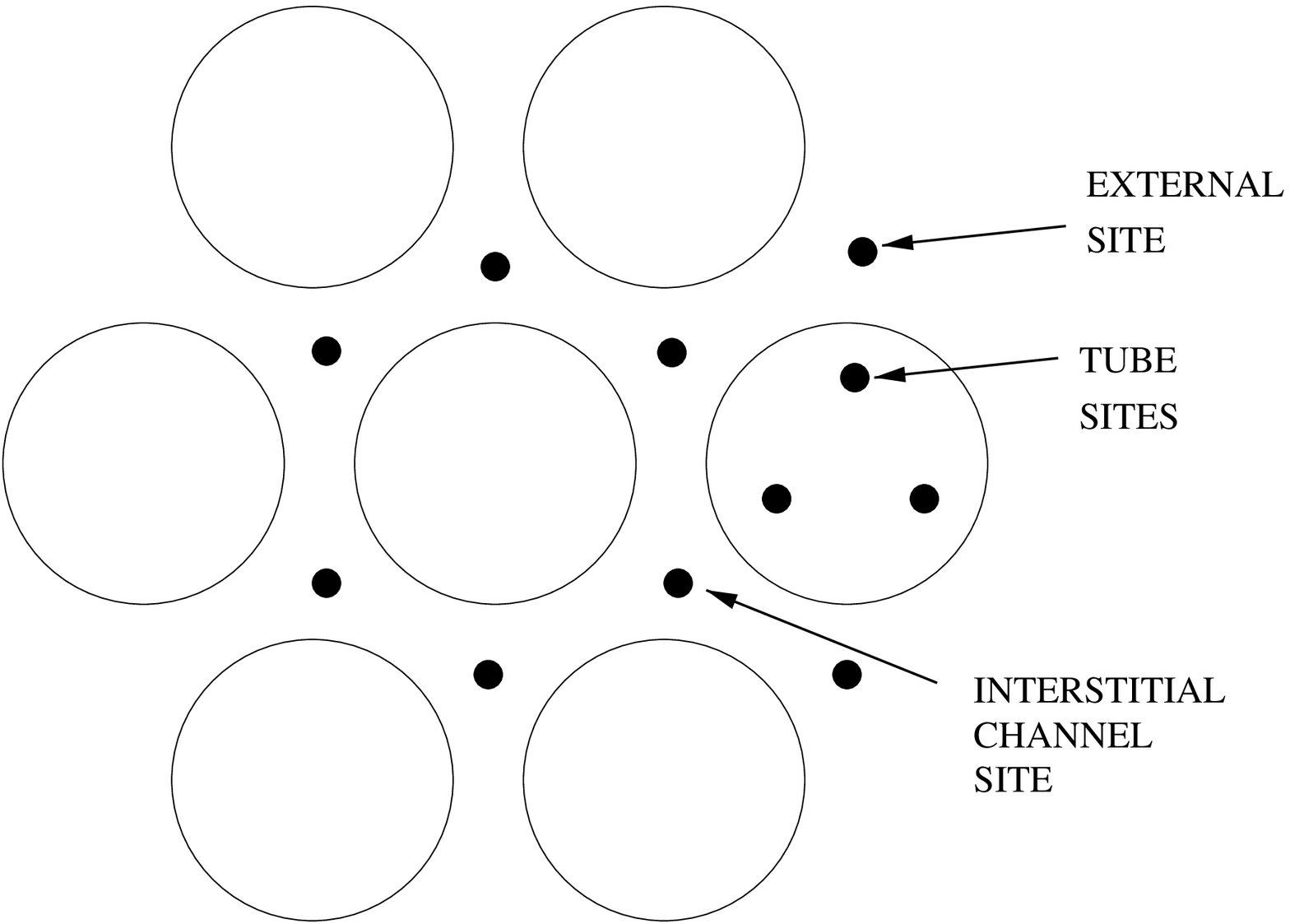}
\end{figure}
\vspace{7cm}
\begin{center}
{\bf FIG. \ref{fig:sites}}
\end{center}

\newpage
\begin{figure}[ht]
\epsfysize=5.in \epsfbox{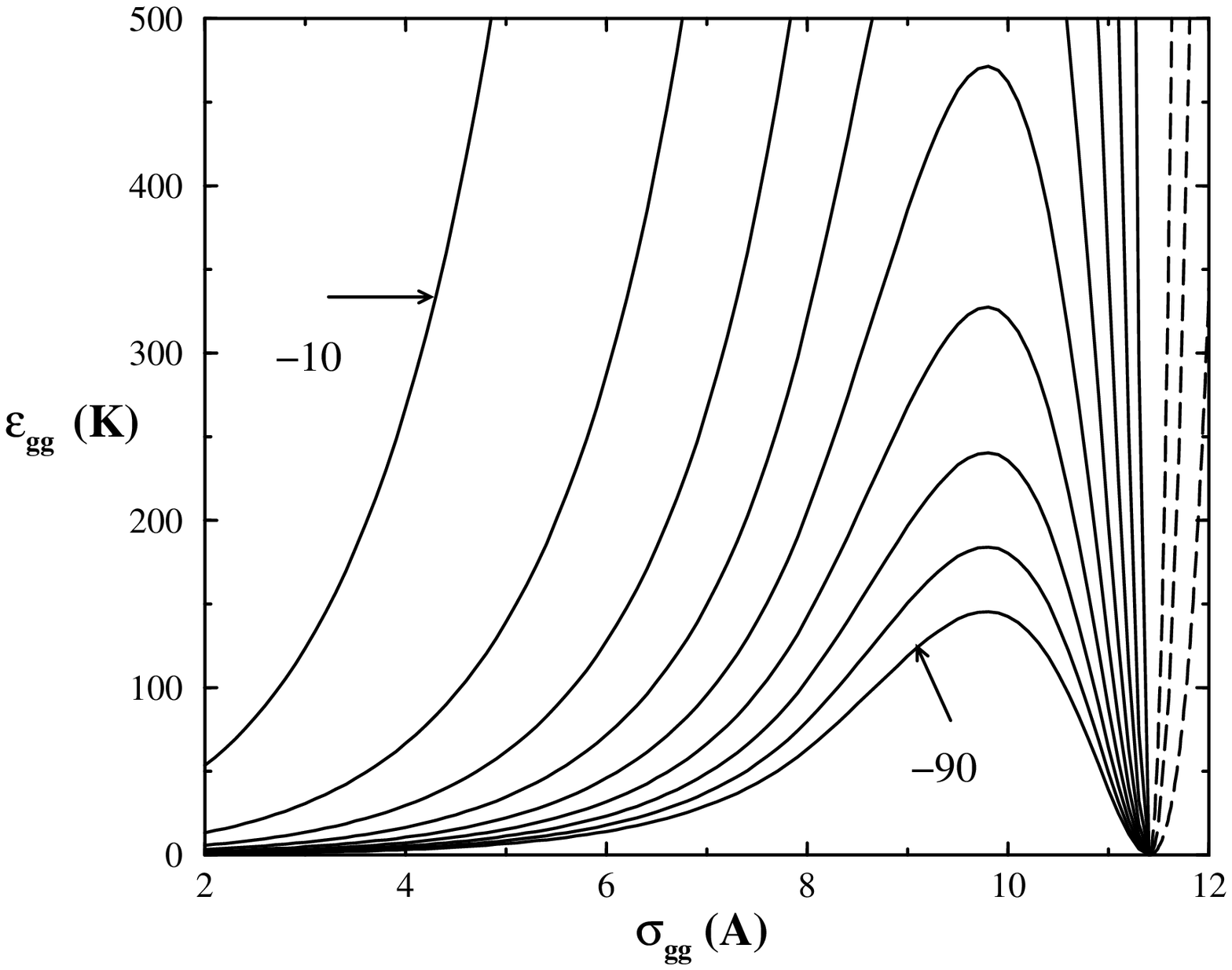}
\end{figure}
\vspace{7cm}
\begin{center}
{\bf FIG. \ref{fig:NT_contour}}
\end{center}

\newpage
\begin{figure}[ht]
\epsfysize=5.in \epsfbox{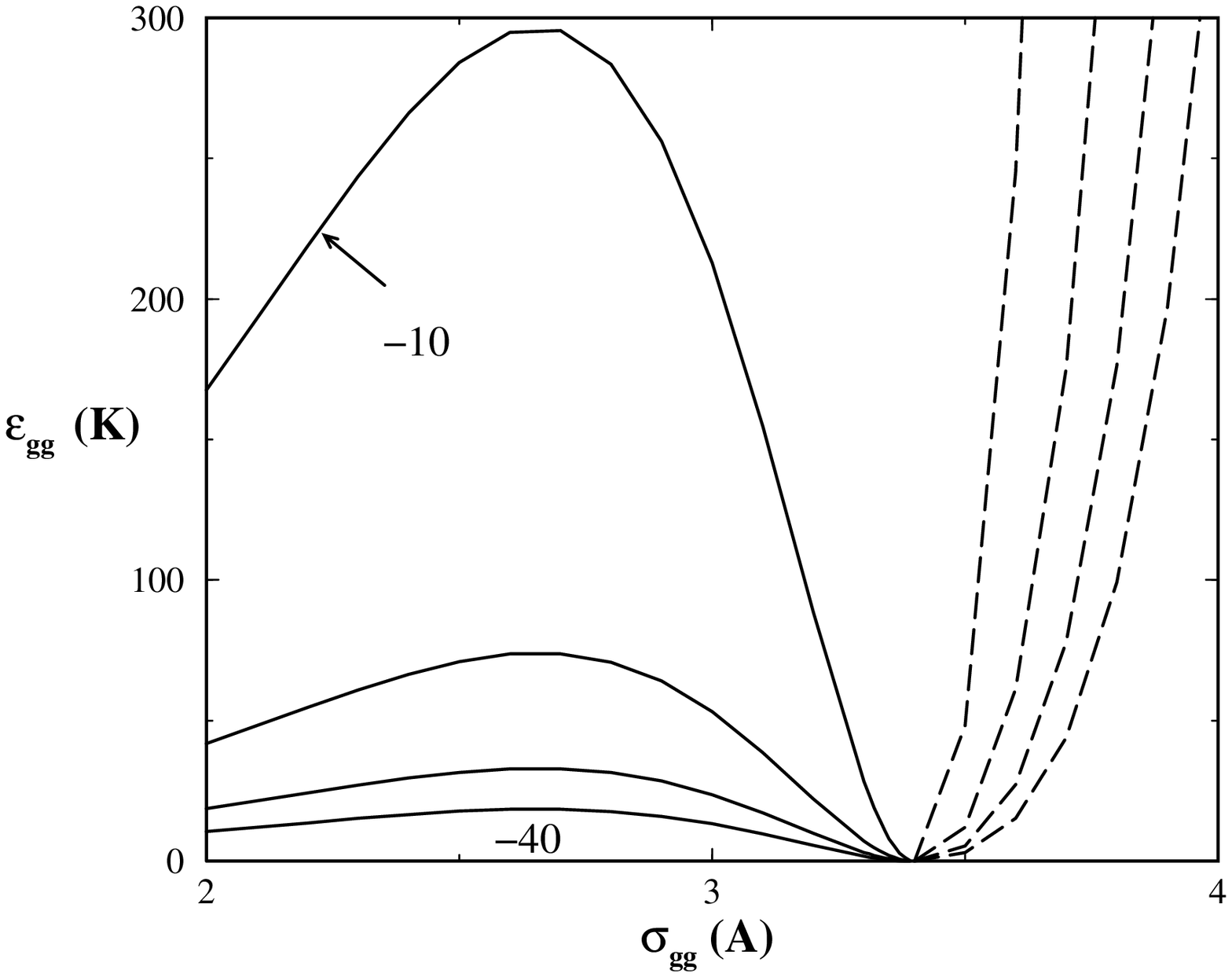}
\end{figure}
\vspace{7cm}
\begin{center}
{\bf FIG. \ref{fig:IC_contour}}
\end{center}

\newpage
\begin{figure}[ht]
\epsfysize=4.6in \epsfbox{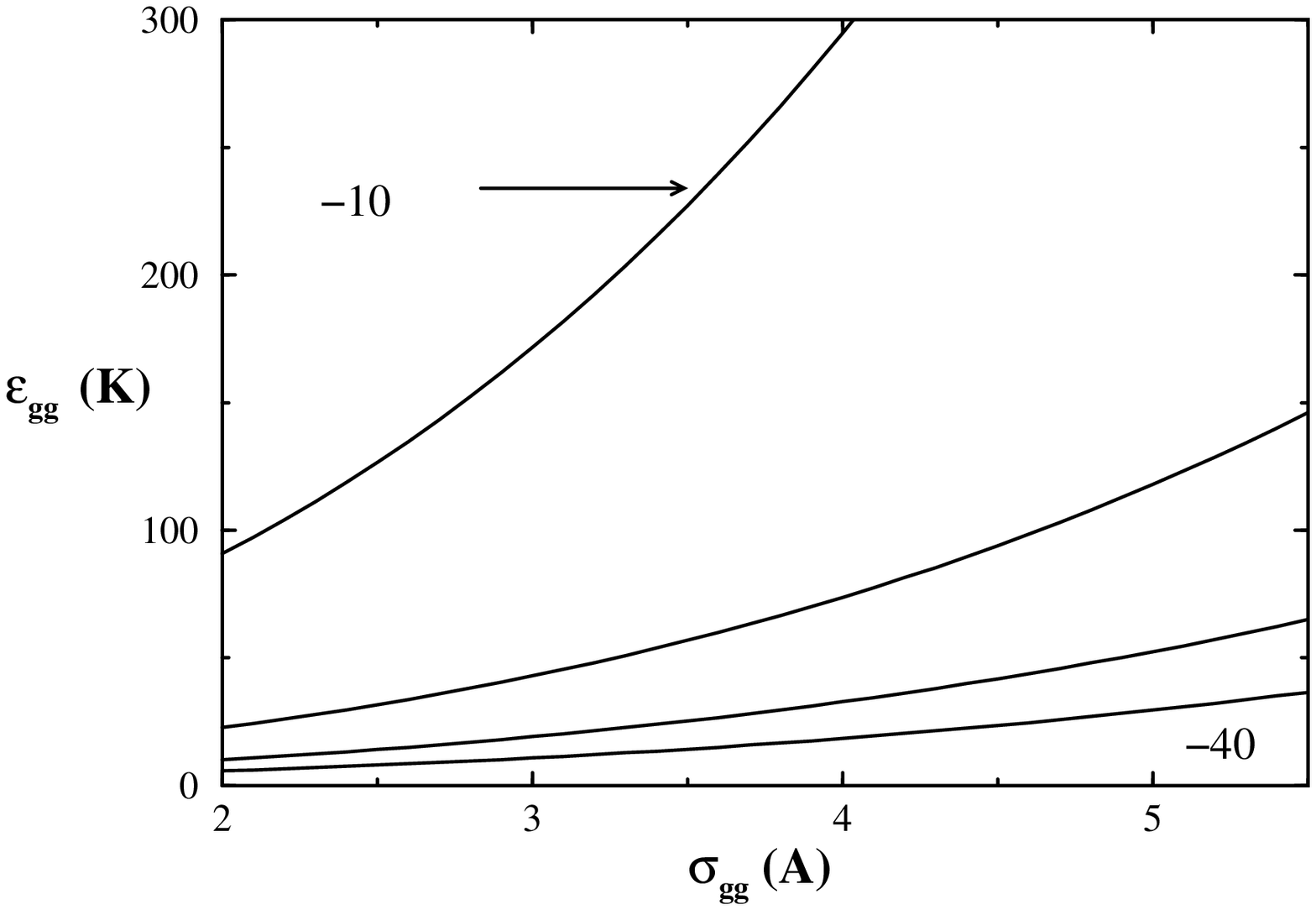}
\end{figure}
\vspace{7cm}
\begin{center}
{\bf FIG. \ref{fig:GR_contour}}
\end{center}

\newpage
\begin{figure}[ht]
\epsfysize=5.in \epsfbox{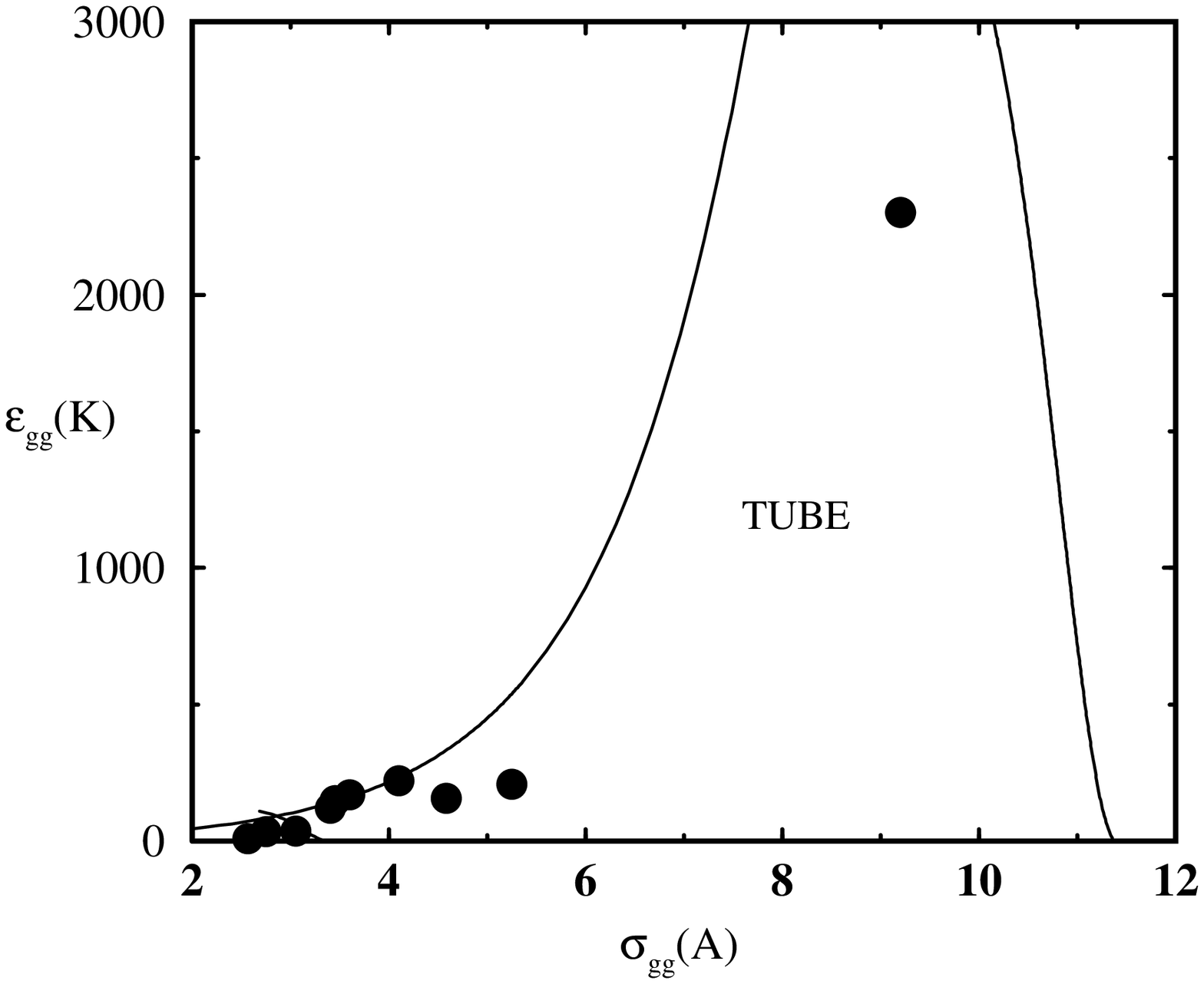}
\end{figure}
\vspace{7cm}
\begin{center}
{\bf FIG. \ref{fig:10.high}}
\end{center}

\newpage
\begin{figure}[ht]
\epsfysize=5.in \epsfbox{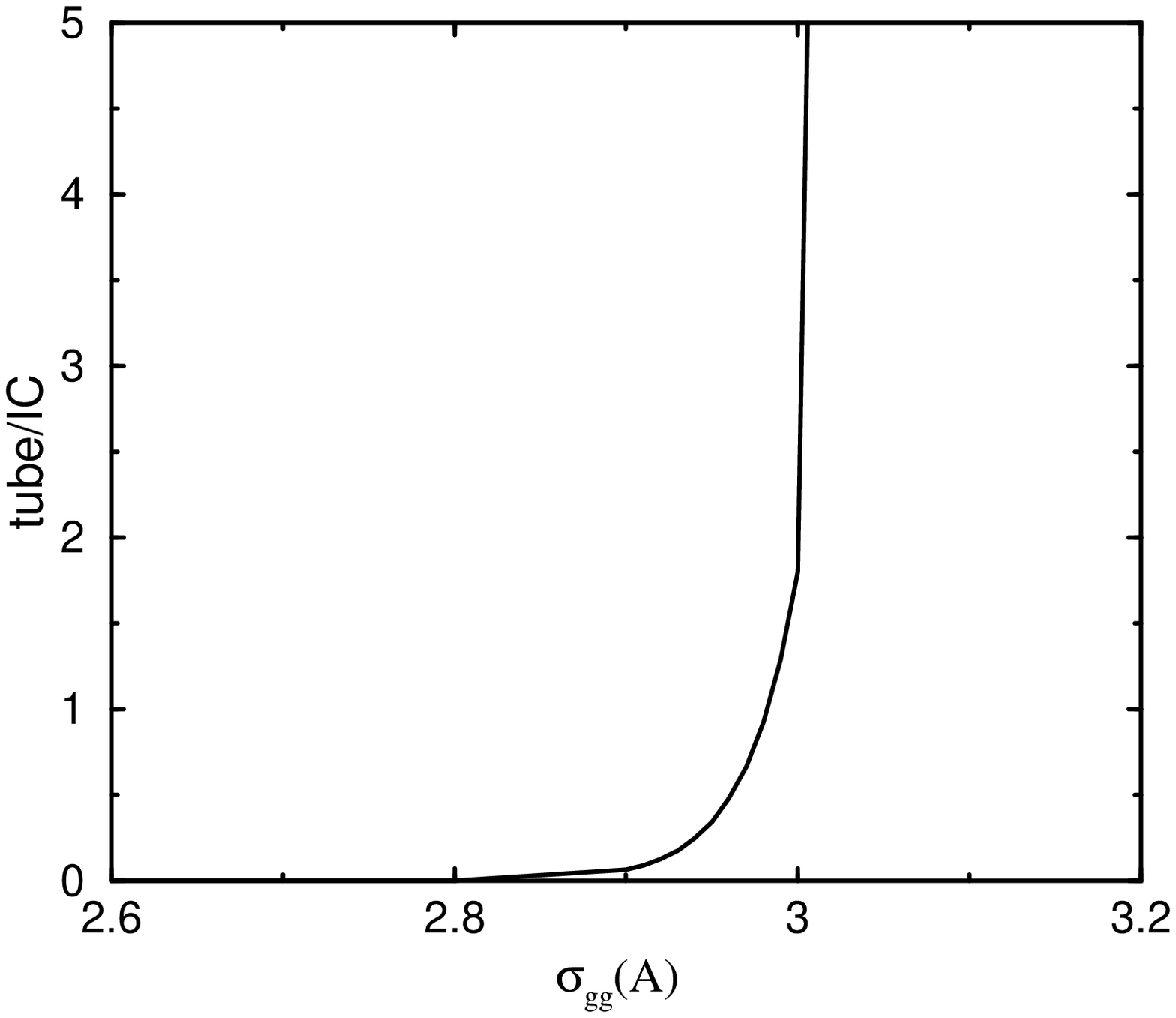}
\end{figure}
\vspace{7cm}
\begin{center}
{\bf FIG. \ref{fig:henry}}
\end{center}

\newpage
\begin{figure}[ht]
\epsfysize=5.in \epsfbox{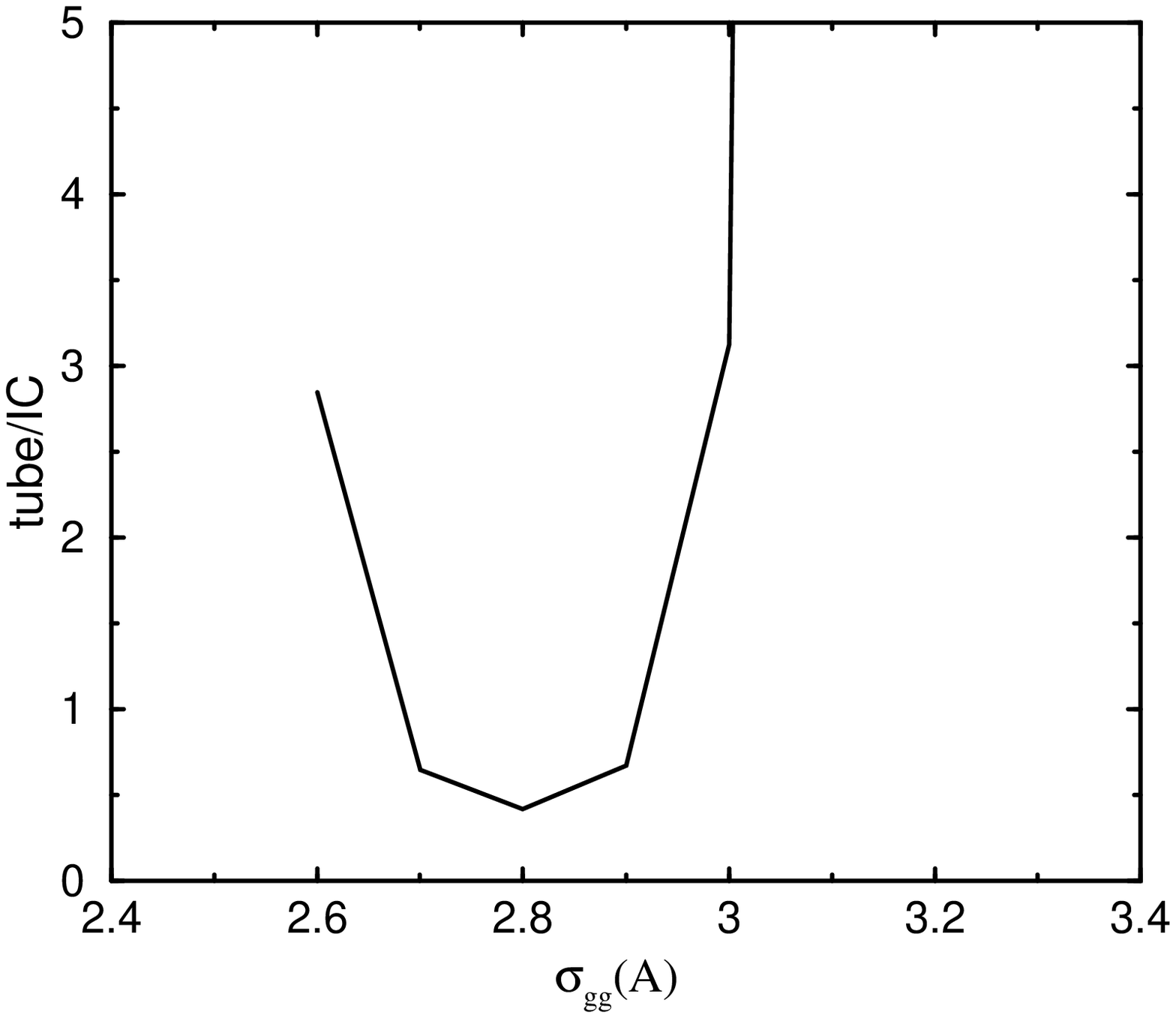}
\end{figure}
\vspace{7cm}
\begin{center}
{\bf FIG. \ref{fig:henry.77}}
\end{center}


\begin{thebibliography}{50}
\bibitem{distance}
As shown in Ref. \cite{stan1}, the equilibrium distance for atoms/molecules 
adsorbed inside a large nanotube ($R > 1.2\ \sigma_{g{\SS C}}$) is slightly 
greater than at a flat surface. In contrast, in small tubes and in IC's this 
distance may be significantly smaller than in the case of a flat surface.
\bibitem{grooves}
For the external surface we have studied the film growth up to high coverage;
we will report these results elsewhere.
\end{thebibliography}
\end{document}